\documentclass[aps,prr,reprint,groupedaddress]{revtex4-2}
\usepackage{amsmath,amssymb,mathtools}
\usepackage{graphicx}
\usepackage{dcolumn}
\usepackage{bm}
\usepackage[hidelinks]{hyperref}
\hypersetup{pdfpagemode=UseNone}
\usepackage{siunitx}
\usepackage{physics2}
\usepackage[italicdiff]{physics}
\usephysicsmodule{braket}

\newcommand{\sSz}{{}^1S_0}
\newcommand{\tSo}{{}^3S_1}
\newcommand{\sPo}{{}^1P_1}

\newcommand{\tPo}{{}^3P_1}
\newcommand{\tPt}{{}^3P_2}

\newcommand{\fermi}{{}^{171}\mathrm{Yb}}

\allowdisplaybreaks[1]

\begin{document}

\preprint{}

\title{Plane-selective manipulations of nuclear spin qubits\\ in a three-dimensional optical tweezer array}


\author{Toshi Kusano}
 \email[]{kusano@yagura.scphys.kyoto-u.ac.jp}
 \affiliation{Department of Physics, Graduate School of Science, Kyoto University, Kyoto 606-8502, Japan}

\author{Yuma Nakamura}
  \affiliation{Department of Physics, Graduate School of Science, Kyoto University, Kyoto 606-8502, Japan}

\author{Rei Yokoyama}
  \affiliation{Department of Physics, Graduate School of Science, Kyoto University, Kyoto 606-8502, Japan}

\author{Naoya Ozawa}
  \affiliation{Department of Physics, Graduate School of Science, Kyoto University, Kyoto 606-8502, Japan}

\author{Kosuke Shibata}
  \affiliation{Department of Physics, Graduate School of Science, Kyoto University, Kyoto 606-8502, Japan}

\author{Tetsushi Takano}
  \affiliation{Department of Physics, Graduate School of Science, Kyoto University, Kyoto 606-8502, Japan}

\author{Yosuke Takasu}
  \affiliation{Department of Physics, Graduate School of Science, Kyoto University, Kyoto 606-8502, Japan}
  
\author{Yoshiro Takahashi}
  \affiliation{Department of Physics, Graduate School of Science, Kyoto University, Kyoto 606-8502, Japan}


\date{\today}

\begin{abstract}
One of the central challenges for a practical fault-tolerant quantum computer is scalability.
A three-dimensional structure of optical tweezer arrays offers the potential for scaling up neutral atom processors.
However, coherent {\it local} operations, essential for quantum error correction, have yet to be explored for this platform.
Here, we demonstrate plane-by-plane initialization of nuclear spin qubits of $\fermi$ atoms in a three-dimensional atom array and plane-dependent coherent temporal evolution of qubits, as well as plane-selective qubit manipulation by exploiting the plane-selective excitation of the atoms from the $\sSz$ to the $\tPt$ state.
This plane-selective manipulation technique paves the way for quantum computing and quantum simulation in three-dimensional multilayer architectures.
\end{abstract}

\maketitle


Neutral atoms in optical tweezer arrays, which allow individual atom control and Rydberg-mediated entanglement generation, have the potential to make a significant contribution to quantum science and technology~\cite{Computer2021,QuantumSimulator2021,kaufmanNi2021}. Recent developments in this platform have enabled a wide range of research in precision measurement, quantum simulation, and quantum computing~\cite{ludlow2019tweezer, Browaeys2020, Morgado2021,saffman2010,saffman2016quantum,henriet2020quantum,Fermionic2023}. The scalability of the system is one of the central issues in quantum science for, e.g., the study of quantum many-body physics, quantum-projection-noise-limited precision measurement, and the implementation of fault-tolerant quantum computation (FTQC)~\cite{beverland2022,Yoshioka2024,Derevianko2011,Ludlow2015,XuQian2024}. 
The neutral atom system offers scalability advantages of minimal couplings between multiple qubits and inherent uniformity of qubit quality.
This enables efficient controllability for a large number of atoms, facilitating the development of state-of-the-art programmable large-scale platforms~\cite{Kim2016,Barredo2016,Endres2016,Ebadi2021,scholl2021,Tao2024,Pause2024,norcia2024,gyger2024,manetsch2024,pichard2024}.

Extending the atom tweezer array platform from a commonly adopted two-dimensional (2D) array configuration to a 3D structure is expected to enhance the scalability in quantum processing. The pioneering works that saw the successful generation of 3D optical tweezer arrays~\cite{Liesener2000, Jennifer2002, Jordan2004, Leach2004, Schonbrun2005, Lee2016, Barredo2018,Schlosser2023} demonstrated important protocols of 3D atom-by-atom assembly~\cite{Sinclair2004, Lee2016, Barredo2018, Schlosser2023, lin2024} and flexible controls of Rydberg interaction in 3D directions~\cite{Barredo2018, Kim2020, Song2021, Kim2022}. 
However, the ability to perform plane-selective coherent manipulations, which is one important ingredient in a 3D atom tweezer array quantum processor, remains to be explored. 
This will provide full 3D controllability when combined with the already established local qubit manipulations and measurements for a 2D array system, such as direct local manipulations by individual Raman beams~\cite{Bluvstein2024}, local off-resonant addressing beams combined with globally irradiated resonant beams~\cite{WurtzOtt2009,Weitenberg2011,LeeJessen2013,WangWeiss2015,YangWeiss2016,Labuhn2014,Sylvain2017,Bornet2024,XiaSaffman2015,Graham2019,GrahamSaffman2022,Levine2018,Omran2019,Barnes2022,Burgers2022,LewisKKNi2024}, and more recent movement-induced shifts using shuttling techniques~\cite{Chen2022,Shaw2024}.

The 3D structure is also beneficial for implementing efficient quantum error-correcting codes such as 3D topological codes~\cite{Dennis2002,Bombin2007_PRB,Bombin2007,Huang2023}, which have favorable features such as the implementation of transversal non-Clifford logical gates~\cite{Bombin2007,Bombín_2015,Kubica2015,Vasmer2019,Bluvstein2024} and single-shot decoding~\cite{Bombin2015,Quintavalle2021,Kubica2022,Bridgeman2024,Stahl2024}. To realize these codes, one promising approach entails shuttling techniques in a real two-dimensional plane~\cite{Bluvstein2022,Bluvstein2024}. Another promising approach is to trap qubits in a real 3D space. In the latter approach, 3D connectivity is achievable by nearest-neighbor interactions, which could generate the entanglement within an order of microseconds, as recently demonstrated in experiments of high-fidelity two-qubit gate operations~\cite{Evered2023,peper2024,tsai2024,cao2024,radnaev2024}.
Moreover, the realization of 3D structure and 3D controllability could be a milestone in the development of four-dimensional codes~\cite{Dennis2002,Alicki2010}, which possess self-correcting properties with coherence times that scale indefinitely with system size~\cite{Terhal2015,Brown2016}. This could be accomplished by incorporating an extra dimension, such as a synthetic dimension~\cite{Ozawa2019} or a moving tweezer array~\cite{Bluvstein2022,Bluvstein2024}, into the existing 3D system.
Note that an optical lattice system is one of the 3D quantum systems with high controllability in which local controls have been reported including coherent manipulations with individual atom addressing beams~\cite{WangWeiss2015, YangWeiss2016} and plane-selective state preparation and imaging using magnetic and electric field gradients~\cite{Schrader2004, Karski_2010, Edge2015, Yamamoto_2016, HanCho2019,Trautmann2023, William2022}. 

\begin{figure*}[tp]
    \centering
    \includegraphics[width=\textwidth]{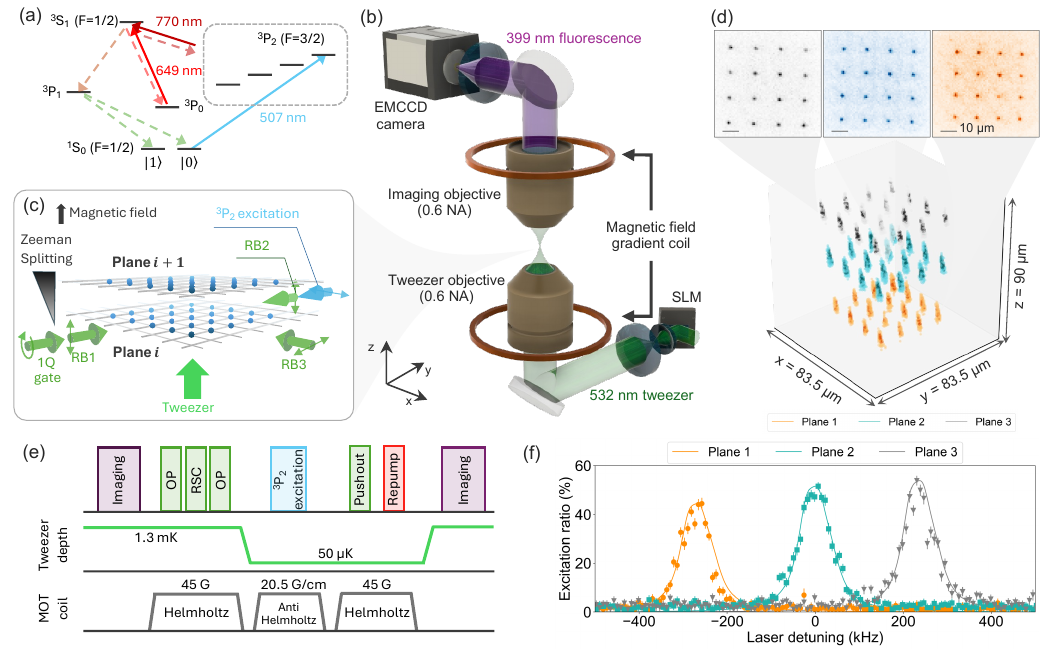}
    \caption{Overview of the 3D ytterbium optical tweezer array. (a) Relevant energy diagram of $\fermi$ atom showing the transitions used to plane-selective controls. (b) Tweezer beam path and imaging system. 3D optical tweezer arrays are generated by combining a quadratic and grating phase hologram displayed on a SLM. The imaging objective is dynamically moved in the $z$-direction using a piezo stage to capture the fluorescence of atoms located in different planes. (c) Schematic illustration of the control beams geometry. All control beams are irradiated globally over the entire array. We use $\sSz$-$\tPo$ transition lasers ($\SI{556}{nm}$) for Raman sideband cooling with three Raman beams (RBs) and a single-qubit gate for nuclear spin qubits in the ground state. For plane-selective operation, we apply a $\SI{507}{nm}$ beam to shelve the atoms in the target plane in the presence of a magnetic field gradient to create a different $\tPt$ resonance for a different plane. (d) Reconstructed average fluorescence image of single $\fermi$ atoms in a 4$\times$4$\times$3 cuboid array, where the site spacing setpoint is $(x,y,z)=(10,10,30)$ $\mu$m. For clarity, the fluorescence image for each plane was colored after the data acquisition. (e) Experimental sequence for plane-selective $\tPt$ excitation. OP represents optical pumping. (f) Excitation spectrum of the 3D array. Solid lines show the simulated spectrum taking into account a residual differential light shift with inhomogeneity of the trap depth and magnetic field fluctuations. The error bars represent the standard error of the mean.
    }
    \label{fig:overview}
\end{figure*}

\begin{figure*}[t]
    \centering
    \includegraphics[width=\textwidth]{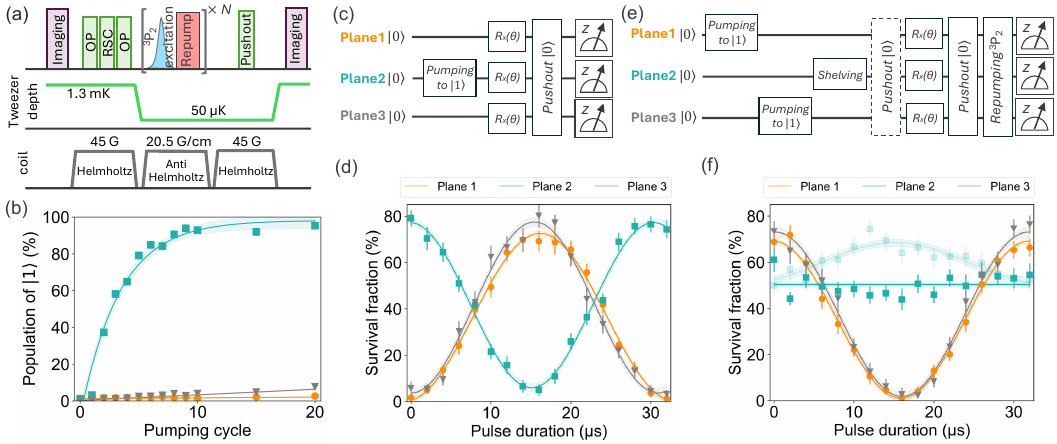}
    \caption{Plane-selective controls. (a) Experimental sequence for plane-selective initialization. After first imaging, we perform optical pumping to $\ket{0}$ and then RSC in 1.3 mK deep tweezer, followed by irradiating the $N$ pulse trains of the $\tPt$ excitation and repumping beams under a magnetic field gradient. (b) Population of $\ket{1}$ after several pumping cycles. The atoms in Plane 2 are selectively pumped to $\ket{1}$ with a fidelity of 95.3(2.0)\% (survival probability corrected). (c)-(f) Plane-selective manipulations. (c), (e) Quantum circuit representations of experimental sequences for the plane-selective Rabi oscillation after (c) selective initialization and (e) selective shelving. (d), (f) Rabi oscillation between $\ket{0}$ and $\ket{1}$ states. (d) We observe the coherent Rabi oscillations of the qubits in Planes 1 and 3, while the qubits in Plane 2 exhibit the bit-flipped behavior after running the (c) circuit. (f) The qubits in Planes 1 and 3 show coherent Rabi oscillations starting from the initially prepared $\ket{1}$ state, after running the (e) circuit. In contrast to (d), the flat line of Plane 2 (blue) indicates noncoupling to the ground state manipulation while the limited shelving fidelity results in a residual oscillation (light blue). The circuit for obtaining blue (light blue) data includes (does not include) the pushout pulse before the nuclear spin control [dashed box in (e)]. In (b), (d), and (f), error bars and shaded regions represent the standard error of the mean and 1$\sigma$-confidence intervals of the fit, respectively.
    }
    \label{fig:pumping}
\end{figure*}

In this work, we report the demonstration of plane-selective manipulations of single ytterbium atoms in a holographically generated 4$\times$4$\times$3 cuboid atom tweezer array. 
We successfully implement local operations in the 3D atom tweezer array using global controls applied to the entire atomic array. 
Specifically, by working with the ground-state nuclear spin qubit of $\fermi$ defined as $\ket{0} = \ket{\sSz, m_F = +1/2}$ and $\ket{1} = \ket{\sSz, m_F = -1/2}$ as well as the magnetic field sensitive metastable state $\tPt$ under a magnetic field gradient [\hyperref[fig:overview]{Fig.~1(a)}], we simultaneously realize both coherent manipulation of the qubit and 
the plane-selectivity in the initialization and shelving.
The demonstrated plane or space-selective excitation to the metastable state under a magnetic field gradient represents a scalable approach, as the addressing spectrum depends solely on the distance between atoms and is independent of the number of qubits.
Our ability to perform plane-selective manipulations is highlighted by the successful demonstration of local coherent qubit rotations $R_X(\theta)$ on specific planes, while on the other selected plane the operations of $R_X(\theta=0)=I$ are performed.
These results underscore the usefulness of an optical-metastable-ground (omg) architecture of $\fermi$ atoms~\cite{Allcock2021,Chen2022} in 3D tweezer arrays, leading to the feasibility of the midcircuit operations~\cite{Lis2023,Ma2023,Scholl2023}, a significant step towards FTQC in 3D tweezer arrays.

{\it 3D Optical Tweezer Array}. 
Our 3D optical tweezer array system utilizes a spatial light modulator (SLM) to generate holograms that include Fresnel lens phases to shift tweezer positions in the $z$-direction [\hyperref[fig:overview]{Fig.~1(b)}]~\cite{Lee2016, Barredo2018}. 
While our experimental setup is basically the same as that in our previous work~\cite{nakamura2024}, we have incorporated a piezo stage (P-528.ZCD, PI) to dynamically move the imaging objective lens, enabling the imaging of the atoms on the individual planes [\hyperref[fig:overview]{Figs.~1(b) and (c)}]. 
This imaging system can typically focus on each plane within approximately 20 ms, which is sufficiently fast for the experiments described in this paper. Each plane is then imaged with a $\SI{60}{ms}$ exposure time.
The 4$\times$4$\times$3 cuboid array structure is successfully imaged in this way, as shown in \hyperref[fig:overview]{Fig.~1(d)}.
All control laser beams are irradiated globally across the entire array [\hyperref[]{Fig.~1(c)}]. 
The magnetic field gradients required for local manipulation experiments are generated by the anti-Helmholtz coil, which also serves as the coil used for magneto-optical trapping (MOT).
The details of the experiments such as the Fresnel lens phase implementation, Raman sideband cooling (RSC), and tweezer homogenization, are described in Sec.~S1 of the Supplemental Material (SM).

{\it Plane-selective Control}. The most essential ingredient for plane-selective control of the 3D array in this work is the spectral addressing using the magnetic-field-sensitive metastable $\tPt$ state under a magnetic field gradient.
The hyperfine manifold $F=3/2$ in the $\tPt$ state has a Zeeman splitting of $\SI{2.5}{MHz/G}\times m_F$, resulting in frequency shifts of $\SI{7.7}{kHz/\micro m}$ for $m_F = 3/2$ when applying a magnetic field gradient of $\SI{20.5}{G/cm}$ in our system. 

As is shown in \hyperref[fig:overview]{Fig.~1(e)}, when performing the $\tPt$ excitation, we decrease the tweezer depth to $\SI{50}{\micro K}$ to reduce the line-broadening effect due to differential light shift (DLS) between the $\sSz$ and $\tPt$ states.
This allows for well-resolved plane-selective excitation to the $\tPt$ state at the current tweezer array spacing, as shown in \hyperref[fig:overview]{Fig.~1(f)}. The observed spectral separation of $\SI{7.76(2)}{kHz/\micro m}$ is close to the designed value of $\SI{7.7}{kHz/\micro m}$.
The solid line in \hyperref[fig:overview]{Fig.~1(f)} represents the simulated spectrum. To simulate the spectrum of the carrier component, we analyze the systematic effects arising from the residual DLS and the Zeeman shift of the $\ket{\tPt, F=3/2,m_F=3/2}$ state. The dominant source of noise in our system is the ripple in the current of the power supply for the magnetic field gradient coil (relative standard deviation of $\SI{0.3}{\%}$). Based on our analysis, the simulated carrier spectrum linewidth is $\SI{53.1}{kHz}$ [full width at half maximum (FWHM)]. This broad linewidth obscures the sideband structure of $\pm\SI{28}{kHz}$, resulting in a total linewidth of $\SI{77}{kHz}$, which reproduces the experimental data well as shown in \hyperref[fig:overview]{Fig.~1(f)}. 

Regarding the $\tPt$ excitation fidelity, with a square pulse irradiation for $\SI{5}{ms}$, we observe the excitation fidelity of $44.9(1.8)\si{\%}$, $53.2(1.7)\si{\%}$, and $51.4(1.8)\si{\%}$ for Planes 1, 2, and 3, respectively.
The limited excitation fraction is attributed to shot-to-shot detuning errors arising from the broad spectral linewidth. To improve the excitation fidelity in subsequent plane-selective control experiments, a hyperbolic secant (HS1) pulse is employed~\cite{Silver1984,Silver1985,Garwood2001,Roos2004} with a typical frequency scan range of $\pm\SI{30}{kHz}$. As a result, the excitation ratios for Planes 1, 2, and 3 are improved to $77.7(1.7)\%$, $83.2(2.1)\%$, and $82.4(2.0)\%$, respectively. Further improvements of the fidelity can be achieved by reducing the current noise and by employing composite pulses~\cite{Lis2023,Muniz2024}.

The achieved plane-selectivity in the spectroscopy is utilized to demonstrate a plane-selective state initialization via the repumping process of the $\tPt$ state [\hyperref[fig:overview]{Fig.~1(a)}].
The pulse sequence is shown in \hyperref[fig:pumping]{Fig.~2(a)}.
First, all atoms in the array are initialized to the $\ket{0}$ state by optical pumping via the $\ket{1}\leftrightarrow \ket{\tPo, F=3/2, m_F=1/2}$ transition. Subsequently, the atoms trapped in a particular plane are selectively excited to the $\ket{\tPt, F=3/2, m_F=3/2}$ state.
The repumping of the atoms in the $\tPt$ state via the $\tSo$, $F=1/2$ state eventually results in a random spontaneous decay from the $\tPo$ state to either $\ket{0}$ or $\ket{1}$ in the ground state. 
The result of successful plane-selective pumping is shown in \hyperref[fig:pumping]{Fig.~2(b)}. 
After 20 pumping cycles, the atoms on Plane 2 are initialized to $\ket{1}$ with a fidelity of 95.3(2.0)\% (survival probability corrected). The finite population of $\ket{1}$ in Planes~1 and 3 is due to the small but non-zero excitation probability to the $\tPt$ state. This crosstalk could be suppressed by narrowing the excitation linewidth or increasing a magnetic field gradient.
Note that, in the current setup, we change the magnetic field from a plane-selective pumping condition where a magnetic field gradient is applied, to a state-selective measurement condition using a pushout beam where a $\SI{45}{G}$ $z$-biased magnetic field is applied. 
This change can cause bit-flips if the magnetic field control is not sufficiently slow, leading to the
remaining initialization error. Unlike experiments with alkali-metal atoms with large magnetic moments, adiabaticity in magnetic controls of nuclear spin qubits should be more careful.
This error can be reduced by performing a state-selective readout, which does not require a strong bias magnetic field for the pushout process~\cite{Lis2023}.

The demonstrated plane-selective initialization is then utilized to further perform plane-dependent
coherent temporal evolution of qubits in a 3D tweezer array.
As described in the quantum circuit of \hyperref[fig:pumping]{Fig.~2(c)}, we initialize the atoms in Plane 2 to $\ket{1}$ selectively, and then apply a nuclear spin control beam to the atoms globally in the direction of the horizontal plane [\hyperref[fig:overview]{Fig.~1(c)}]. 
After the circuit operation, we observe a Rabi oscillation in Plane 2 that is phase-shifted by $\pi$ from other planes [\hyperref[fig:pumping]{Fig.~2(d)}]. 

For the feasibility of midcircuit operations in 3D tweezer arrays, we demonstrate that the atoms in a particular plane can be protected from the ground-state nuclear spin manipulation by the $\tPt$ shelving technique.
The quantum circuit is illustrated in \hyperref[fig:pumping]{Fig.~2(e)}. To suppress the influence of the pushout beam [dashed box in \hyperref[fig:pumping]{Fig.~2(e)}] prior to the nuclear spin control, Planes~1 and 3 are initialized to the $\ket{1}$ state with 10 pumping cycles at the beginning of the circuit. Subsequently, we shelve atoms in Plane~2 with the HS1 pulse, followed by pushing out the $\ket{0}$ state before applying the nuclear spin manipulation pulse to the entire array. \hyperref[fig:pumping]{Figure~2(f)} shows the measurement results after this quantum circuit. While atoms in Planes~1 and 3 exhibit Rabi oscillations (orange and gray), shelved atoms in Plane~2 are independent of the pulse width (blue), indicating that the atoms in Plane~2 are selectively decoupled from the ground-state manipulation. The light blue data in \hyperref[fig:pumping]{Fig.~2(f)} show the measurement result by a sequence without the pushout beam [dashed box in \hyperref[fig:pumping]{Fig.~2(e)}], and a residual oscillation of the ground state nuclear spin qubit is observed. To quantify the residual oscillation, we fit the data by a function $P_sP_{3\mathrm{P}2}P_r + P_s(1-P_{3\mathrm{P}2})(1+\sin\qty(\Omega t + \phi))/2$ with the $\tPt$ excitation fidelity $P_{3\mathrm{P}2}$, the Rabi frequency $\Omega$, and phase $\phi$ as free parameters. Here, the survival probability $P_s$ is determined from the maximum value of the data $74.4(3.5)\si{\%}$, and the repumping fidelity $P_r$ is $98.2(4.2)\si{\%}$, which is obtained as described in Sec.~S3 in the SM. From the fitting, the $\tPt$ excitation fraction is determined to be $72.0(2.5)\si{\%}$, which is lower than $83.2(2.1)\si{\%}$ obtained in the plane-selective excitation experiment. We attribute this decreased fidelity to the instability of the excitation laser frequency after switching the laser frequency to the resonance of each plane, where the $\tPt$ excitation laser frequency is currently tuned by switching the locking frequency to an ultra-low-expansion (ULE) cavity used for the laser frequency stabilization. This can be solved by switching the frequency using a conventional double-path acousto-optic modulator.

{\it Discussion}. A shorter interplane distance is desirable for achieving sufficiently strong Rydberg interactions in
a 3D structure. In our current experiment, the spacing smaller than  $\SI{30}{\micro m}$ does not provide sufficient spectral resolution to address different layers. This limitation arises from the broadened linewidth of the $\ket{0} \leftrightarrow \ket{\tPt, F=3/2, m_{F}=3/2}$ transition of $\fermi$ atoms due to the magnetic field fluctuation induced by the gradient coils.
The effect of the coil current noise is actually quantitatively evaluated from the difference in the measurement of the excitation linewidth of $\SI{2.6(4)}{kHz}$ (FWHM) for $\ket{\sSz, m_F=1/2} \leftrightarrow \ket{\tPt, F=3/2, m_{F}=3/2}$ in 2D arrays without the application of a magnetic field gradient.
We expect that a straightforward solution of working with a larger magnetic field gradient of $\SI{300}{G/cm}$ as well as suppressing the magnetic field fluctuation by a factor of 50 will enable the experiment to be carried out at a shorter interplane distance (see Sec. S3 in the SM for details). 

A key challenge for quantum computation in 3D tweezer arrays is a plane-selective midcircuit measurement. While the $\sSz \leftrightarrow \sPo$ probe light illuminates the entire array during imaging, only the atomic fluorescence from a single plane is focused onto the camera. Thus, atoms in other planes that are out of focus experience excess scattering. To address this, one promising strategy is to shelve all the atoms that are in nontarget planes into a metastable state, isolating them from the lasers for imaging and cooling. 
In our current experiment, an optical tweezer at a $\SI{532}{nm}$ wavelength and $\SI{1.3}{mK}$ depth is utilized during imaging, causing a severe decrease in survival probability of atoms in the $\tPt$ state, due to the atomic loss by the ionization of the $\tPt$ state (see Sec.~S3 in the SM). Improving the cooling performance during imaging, such as by cooling atoms in tweezers at a magic condition~\cite{Ma2023, Lis2023}, will allow us to image the atoms in shallower depth of the tweezers~\cite{nakamura2024}, facilitating
plane-selective midcircuit measurements with sufficiently high imaging fidelity.

{\it Summary}. We have successfully developed a programmable 3D atom tweezer array of $\fermi$ with the capability of plane-selective manipulation. 
We demonstrate plane-by-plane initialization of the nuclear spin qubits and plane-dependent coherent temporal evolution of qubits, as well as plane-selective qubit manipulation by exploiting the plane-selective excitation of the atoms from the $\sSz$ to the $\tPt$ state under a magnetic field gradient.
While plane-selective imaging is widely used in optical lattice platforms, our work represents the first realization of plane-selective initialization and plane-selective manipulations in an optical tweezer array within a magnetic field gradient.
Furthermore, our plane-selective manipulation technique utilizing the metastable $\tPt$ manifolds enables local control of even magnetically insensitive qubits, showing a sensitivity of $\SI{7.7}{kHz/\micro m}$ in a 20.5~G/cm gradient. This contrasts with the inherent sensitivity of $\SI{1.5}{Hz/\micro m}$ for $\fermi$ nuclear spin qubits in the same magnetic field gradient magnitude.

In addition to quantum computing, a system of 3D atom tweezer arrays with plane-selective coherent controllability also opens up new horizons for quantum simulations. The pyrochlore lattice, a natural platform for quantum spin ice, can be realized by trapping atoms in arbitrary geometries and tuning the parameters of the transverse Ising-like Hamiltonian~\cite{Barredo2018,Shah2024, Semeghini2021}. Additionally, a recent proposal suggests utilizing interspecies (interisotopes) interactions in a 3D tweezer system to generate the ground state of the X-cube model, highlighting the potential of 3D dual-species (dual-isotope) arrays for observing Fracton order~\cite{nevidomskyy2024}.

\begin{acknowledgments}
We thank Keito Saito, Koichiro Higashi and Toshihiko Shimasaki for the experimental assistance and helpful feedback on the manuscript.
This work was supported by Grants-in-Aid for Scientific Research of JSPS (No.\ JP17H06138, No.\ JP18H05405, No.\ JP18H05228, No.\ JP21H01014, No.\ JP22K20356), JST PRESTO (No.\ JPMJPR23F5), JST CREST (No.\ JPMJCR1673 and No. JPMJCR23I3), MEXT Quantum Leap Flagship Program (MEXT Q-LEAP) Grant No.\ JPMXS0118069021, JST Moon-shot R\&D (Grants No.\ JPMJMS2268 and No. JPMJMS2269), and JST ASPIRE (No. JPMJAP24C2).
T.K. acknowledges support from the JST SPRING, Grant No. JPMJSP2110.
Y.N.\ acknowledges support from the JSPS (KAKENHI Grant No.\ 22KJ1949). N.O. acknowledges support from the JSPS (KAKENHI Grant No.\ 24KJ0120).
\end{acknowledgments}

\bibliography{Refs}

\end{document}


\preprint{}

\title{Supplementary Information for ``Plane-selective manipulations of nuclear spin qubits in a three-dimensional optical tweezer array"}





\maketitle


\section{S.1 Methods}
\textbf{\textit{Hologram calculation.}} {\bf \textemdash} The optical tweezer arrays are generated by a spatial light modulator (SLM; X15213-L16, Hamamatsu). The elementary phase generating a tweezer site (of index $m$) at the position $(x_m,y_m,z_m)$ in the chamber is given by the following relation:
    \begin{equation}
        \label{eq:elem_phase}
        \Delta_m(x_s, y_s) = \frac{2\pi M}{\lambda f_{obj}}\qty(x_mx_s + y_my_s) + \frac{\pi M^2 z_m}{\lambda f_{obj}^2}\qty(x_s^2 + y_s^2), \tag{S1}
    \end{equation}
where $(x_s, y_s)$ is the $x$ and $y$ position of the SLM pixel (of index $s$), the magnification of the $4f$-system between the SLM and the objective lens $M=5/3$, the focal length of the objective lens (Special Optics) $f_{obj}=\SI{20}{mm}$ and the wavelength of the tweezer laser (Verdi V-10, Coherent) $\lambda = \SI{532}{nm}$. The first term in Eq.~(S1) represents the grating shift in the horizontal plane that is widely used in 2D holographic tweezer arrays, and the second term represents the Fresnel phase that shifts the focus position in the $z$-direction. The procedure of generating this 3D array from the calculated elementary phases is the same as that in the case of generating a 2D array~\cite{DiLeonardo:07}. 

\textbf{\textit{Atom loading.}} {\bf \textemdash} The experimental sequence is essentially the same as our previous work~\cite{nakamura2024}. We begin with a magneto-optical trap (MOT) of $\fermi$ atoms on the $\sSz - \tPo$ transition ($\SI{556}{nm}$). After a $\SI{0.5}{s}$ MOT loading period, atoms are loaded into a 3D optical tweezer array by increasing the magnetic field gradient from $\SI{10.4}{G/cm}$ to $\SI{17.4}{G/cm}$. We then irradiate red-detuned light-assisted collision (LAC) beams to prepare single atoms in the tweezer array (Fig.~S1(a)).

\textbf{\textit{Atom imaging.}} {\bf \textemdash} The single atoms in 3D tweezer arrays are imaged on the $\sSz - \sPo$ transition ($\SI{399}{nm}$) while simultaneously cooling with MOT beams. The fluorescence from atoms are collected by an objective lens with numerical aperture (NA) of 0.6 and subsequently focused onto an EMCCD camera (iXon-Ultra-897, Andor). In particular, to focus the fluorescence from each plane, we move the objective lens by a piezo stage (P-528.ZCD, PI) to an appropriate position with the typical moving time of the piezo stage of $\SI{20}{ms}$. 
Imaging is performed sequentially on Planes 1, 2, and 3, as illustrated in the experimental sequence shown in Fig.~S1(b). 

In the 4$\times$4$\times$3 cuboid array experiment, the exposure time for each plane is $\SI{60}{ms}$, and the survival probabilities after imaging are typically $\SI{89}{\%}$, $\SI{91}{\%}$ and $\SI{92}{\%}$ for Planes 1, 2, and 3, respectively. These survival probabilities are lower than our typical survival probability of $\SI{96}{\%}$ for 2D arrays due to excess heating of other planes during the measurement of one plane, as the $\SI{399}{nm}$ probe light is illuminated for the entire array while only one plane can be imaged at a time. This issue could be resolved by (1) plane-selective measurement, where planes other than the target plane are shelved to a metastable state, and only the target plane is imaged, or (2) simultaneous imaging of multiple planes using a SLM~\cite{Sun21}. Additionally, the behavior of increasing survival probability from Plane 1 to Plane 3 is attributed to the non-collimated z-direction MOT beam used for cooling during imaging, resulting in varying cooling efficiency for different plane positions. We expect that this variation in the survival probability among planes can be suppressed by improving our optical system.

To acquire the 3D tweezer image in Fig.~1, we first take 60 fluorescence images for each plane and obtain the average image. The piezo stage is then moved by $\SI{1}{\micro m}$, getting another averaged images of 60 shots. This process is repeated until 2D image data covering a range of $\SI{90}{\micro m}$ along the $z$-direction are obtained, which are then reconstructed into a 3D image. While we set the site separation in the $xy$-plane as $\SI{10}{\micro m}$ and the inter-plane distance as $\SI{30}{\micro m}$, the image data show that the $xy$-plane site spacing is $\SI{12.9}{\micro m}$, $\SI{13.8}{\micro m}$ and $\SI{14.6}{\micro m}$ for Planes 1, 2, and 3, respectively (which is calculated by the EMCCD pixel size and the imaging magnification), and the inter-plane distances between Plane 1$-$2, and Plane 2$-$3 are measured to be $\SI{34}{\micro m}$ and $\SI{30}{\micro m}$, respectively. This discrepancy is likely due to the optical systems for tweezer and imaging not being perfect $4f$-systems. However, this mismatch between the setpoint and the actual tweezer positions is not fatal. The positional error can be corrected by calibrating the SLM settings.

    \begin{figure}[htbp]
        \centering
        \includegraphics[width=\linewidth]{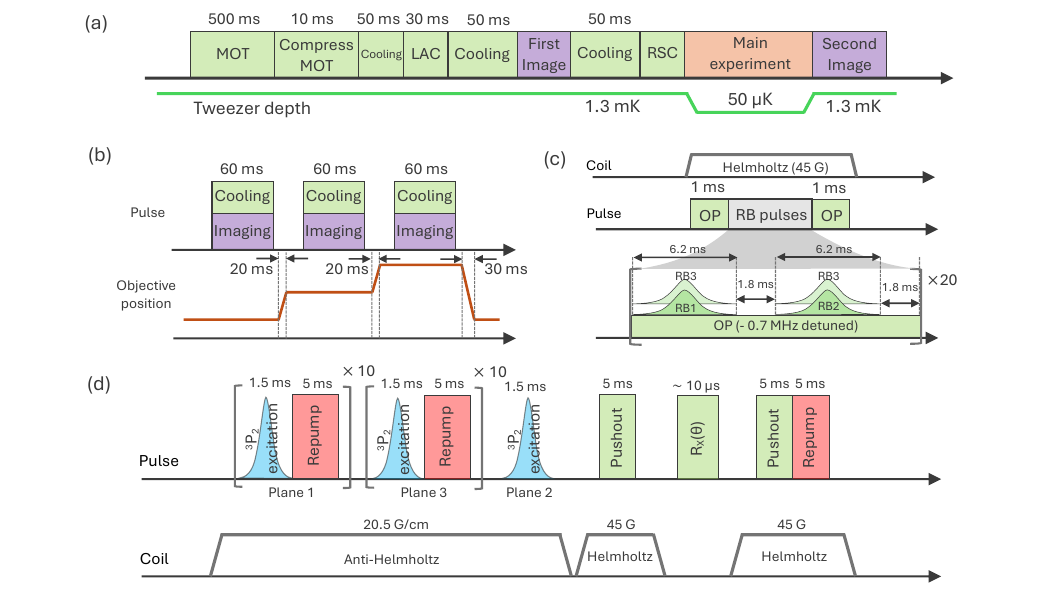}
        \caption{Experimental sequences. (a) Overview of the experimental sequence. Initially, $\fermi$ atoms in a magneto-optical trap (MOT) are compressed to the positions of the tweezer spots. Subsequently, single $\fermi$ atoms are probabilistically loaded into each tweezer spot by irradiating light-assisted collision (LAC) beams. The atomic fluorescence is measured twice, and the survival probability is determined by comparing the results of the two measurements. Between these two measurements, plane-selective manipulations are performed. (b) Imaging sequence. We image the trapped atoms on the $\sSz-\sPo$ transition while simultaneously cooling the atoms on the $\sSz-\tPo$ transition. The objective lens is moved along the $z$-axis with a piezo stage to collect fluorescence from each plane. (c) Raman sideband cooling (RSC) sequence. To initialize the qubit state to the $\ket{0}$ state before and after the RSC, we perform optical pumping (OP) on the $\sSz-\tPo$ transition. The RSC is performed by irradiating multiple Raman beams (RB) from three horizontal directions while simultaneously continuously applying the OP beam. (d) Pulse sequence for plane-selective control experiment of Fig.~2(e). A ten-pulse sequence of $\tPt$ excitation and repumping pulses is used for plane-selective state initialization. The initial pulse sequence and the subsequent sequence perform plane-selective state initializations on atoms in Plane~1 and Plane~3, respectively. A single $\tPt$ excitation pulse is then applied to shelve atoms in Plane 2 to the $\tPt$ state, followed by a manipulation pulse $R_x(\theta)$ for the nuclear spin qubit in the ground state applied to the entire array for plane-selective single-qubit manipulation.}
        \label{fig:sequence}
    \end{figure}

\textbf{\textit{Correcting aberrations.}} {\bf \textemdash} Regarding the correction of aberrations, we utilize the phase of the Zernike polynomials that is widely applied in 2D tweezer systems~\cite{Gordon1997,Nogrette2014}. We optimize the Zernike parameters by maximizing the trap frequency in a 2D array at $z= 0$, where the trap frequency is measured by Raman sideband spectroscopy of the ground state of $\fermi$. The resulting beam waist $w_0$ at z=0 is approximately $\SI{550}{nm}$.

\textbf{\textit{Trap depth homogenization.}} {\bf \textemdash} Homogeneous trap depth of tweezer arrays is essential for efficient quantum control. We achieve the homogenization from the information obtained by performing spectroscopy of the differential light shifts (DLS) of the $\ket{\sSz,m_F=1/2} \leftrightarrow \ket{\tPo,F=3/2,m_F=\pm3/2}$ transition for each atom. This spectroscopy is conducted in a single experimental sequence, starting from MOT loading and ending with the measurement of the survival probability of atoms, by moving the objective lens position. Subsequently, the laser powers for each site are adjusted according to the rule of Eq.~(3) in Ref.~\cite{Nogrette2014}. After several cycles of spectroscopy and power optimization, we achieve $0.5 {\%}$ inhomogeneity of the DLS of the $\ket{\sSz, m_F=1/2}\leftrightarrow\ket{\tPo, F=3/2, m_F=\pm3/2}$ transition for the 4$\times$4$\times$3 cuboid array.

\textbf{\textit{Qubit initialization and readout.}} {\bf \textemdash} The nuclear spin qubit is encoded in the ground state of $\fermi$ as $\ket{0} = \ket{\sSz, m_F = +1/2}$ and $\ket{1} = \ket{\sSz, m_F = -1/2}$.
Qubits are initialized to the $\ket{0}$ state using optical pumping (OP) on the $\sSz - \tPo$ transition. In this initialization process, we irradiate the $\ket{1} \leftrightarrow \ket{\tPo, F=3/2, m_F=1/2}$ resonant beam horizontally for $\SI{1}{ms}$ under a $\SI{45}{G}$ z-bias magnetic field.
State-selective readout is performed by imaging the ground state after selectively removing the atoms in the $\ket{0}$ state. A pushout beam resonant with the $\ket{0}\leftrightarrow\ket{\tPo,F=3/2,m_F=3/2}$ transition is irradiated for $\SI{5}{ms}$ from the same path as the OP beam. To enhance the efficiency of the atom removal, we ramp down the tweezer depth to $\SI{50}{\micro K}$ during the pushout beam irradiation, with a $\SI{45}{G}$ z-bias magnetic field applied. 

\textbf{\textit{Raman sideband cooling.}} {\bf \textemdash} To mitigate the excitation error caused by motional-state-dependent coupling between the ground and metastable states~\cite{takamoto2009,Jenkins2022}, we perform Raman sideband cooling (RSC)~\cite{Kaufman2012,Thompson2013,Jenkins2022} in the horizontal plane prior to plane-selective excitation to the $\tPt$ state (Figs.~S1(a) and (c)). As shown in Fig.~1(b), the three Raman beams (RB) with a detuning of $\sim - \SI{900}{MHz}$ from the $F = 1/2$ of $\tPo$ resonance are irradiated horizontally. The intensities of each RB are set such that the carrier Raman Rabi frequency was $2\pi\times\SI{11.7(2)}{kHz}$ for RB1-3 and $2\pi\times\SI{10.5(2)}{kHz}$ for RB2-3. Despite the trap depth inhomogeneity of $\SI{0.5}{\%}$ as mentioned before, the trap frequencies of the 4$\times$4$\times$3 cuboid array with an interplane distance of $\SI{30}{\micro m}$ are $2\pi\times\SI{176(4)}{\kHz}$, $2\pi\times\SI{176(6)}{\kHz}$, and $2\pi\times\SI{160(5)}{\kHz}$ for Planes 1, 2, and 3, respectively. To suppress the detuning errors in the Raman transition caused by the variations in trap frequency, a hyperbolic secant (HS1) pulse~\cite{Silver1984,Silver1985,Garwood2001,Roos2004} is employed for the RB pulse, where we sweep the frequency by $\SI{70}{kHz}$ for RB1-3 and $\SI{60}{kHz}$ for RB2-3 over $\SI{6.2}{ms}$ to ensure adiabaticity of the transition with the HS1 pulse. The OP during irradiating RB pulses is detuned by $-\SI{0.7}{MHz}$ from the OP frequency used for the qubit initialization to suppress atomic heating.

This results in the motional ground state population of $P_0 = \SI{63(6)}{\%}$ ($\bar{n}_{xy} = 0.59(15)$) in a 3D array with an interplane distance of $\SI{30}{\micro m}$. The limited cooling performance can be attributed to variations in trap frequency among planes due to the variations in beam waist along the $z$ position (Fig.~S3). In fact, the measurement of the motional ground state population in different tweezer structures with a smaller size along the $z$-direction, such as a 4$\times$4$\times3$ cuboid array with $\SI{10}{\micro m}$ interplane distance and a single layer of $7\times7$ 2D square array, show the higher values of $P_0 = \SI{85(6)}{\%}$ ($\bar{n}_{xy} = 0.17(8)$) and $\SI{84(2)}{\%}$ ($\bar{n}_{xy} = 0.19(3)$), respectively, suggesting that achieving uniform cooling across the entire array becomes challenging when a 3D array is arranged over a wide area along the $z$-direction. We anticipate that future improvements in shape homogenization using apodization techniques~\cite{chew2024} or erasure cooling methods~\cite{shaw2024_erasure} will enhance the cooling performance.

\textbf{\textit{Nuclear spin qubit control}} {\bf \textemdash}
The single-qubit gate beam is generated using the same laser source as that used for Raman sideband cooling, which is detuned by -900~MHz from the $\tPo, F=1/2$ resonance. This beam is irradiated from the same direction as the RB1 beam (Fig.~1(c)). During gate operation, a bias magnetic field of 0.9 G is applied in the $z$-direction. The operation is performed using a strong-driving regime with a Rabi frequency much larger than the Zeeman splitting of 0.7 kHz~\cite{Jenkins2022}. The Rabi frequencies for Planes 1, 2, and 3 are $2\pi\times \SI{29.7(8)}{kHz}$, $2\pi\times \SI{32.8(9)}{kHz}$, and $2\pi\times 32.4(1.1)~\si{kHz}$, respectively. The variation in Rabi frequencies across the planes is due to the non-uniform power distribution resulting from the $\SI{156}{\micro m}$ beam waist of the single-qubit gate beam in the $z$-direction.

To characterize the single-qubit gate fidelity, we evaluate the average single-Clifford gate fidelity using randomized benchmarking~\cite{Knill2008, Magesan2011}. We randomly sample gates from the single-Clifford gate group and then compile the sampled Clifford gates into sequences of $X(\pi/2)$ and $Z(\pi/2)$ pulses. On average, there are 3.5 of $X(\pi/2)$ and $Z(\pi/2)$ pulses per single Clifford gate, achieving performance comparable to that of previous work~\cite{Jenkins2022}. Here, the $Z$ gate beam is output from a different laser source than that used for the $X$ gate, detuned by -860~MHz from the $\tPo, F=1/2$ resonance, and is irradiated onto the atoms from the same direction as the RB3 beam (not shown in Fig.~1(c)). In the randomized benchmarking experiment, a bias magnetic field of 0.9~G is applied in the direction aligned with the $Z$ gate beam. We note that this randomized benchmarking experiment is performed in a 6$\times$6 two-dimensional array with $\SI{6}{\micro m}$ site spacing to evaluate the gate error excluding the power inhomogeneity of the gate beam. Figure~S2 shows the results of randomized benchmarking when the Rabi frequencies of the $X$ and $Z$ gate beams are $2\pi\times \SI{9.6(2)}{kHz}$ and $2\pi\times \SI{81.8(6)}{kHz}$, respectively. From the fitting of $P(m) = A_0(p^m + 0.5)$, we obtain the average gate fidelity $\mathcal{F} = (p+1)/2 = \SI{99.71(2)}{\%}$, where $m$ is the gate depth, and $p$ and $A_0$ are fitting parameters.

    \begin{figure}[htbp]
        \centering
        \includegraphics[width=0.6\linewidth]{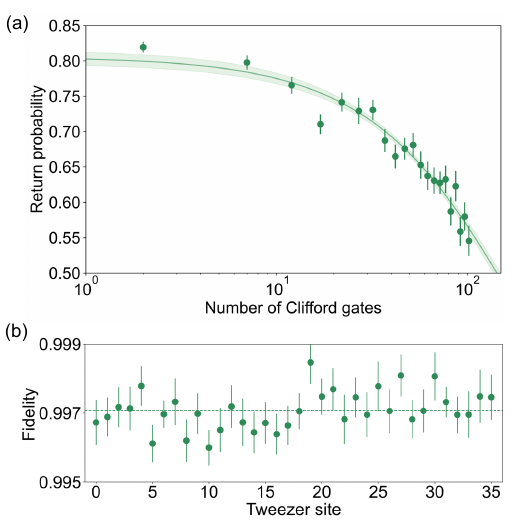}
        \caption{Nuclear spin qubit control characterization. (a) Clifford randomized benchmarking decay, averaged across 64 sites. We obtain an averaged Clifford gate fidelity of $\mathcal{F}=\SI{99.71(2)}{\%}$. At each depth, 150 different random circuits are sampled. The error bar represents the standard error over the 150 sets of random circuits. The shaded area is the $1\sigma$-confidence intervals of the fit. (b) Site-resolved analysis of averaged Clifford gate fidelities. }
        \label{fig:RB}
    \end{figure}

\textbf{\textit{Excitation to the metastable state.}} {\bf \textemdash}
The excitation laser for the $\sSz-\tPt$ transition ($\SI{507}{nm}$) is generated by second harmonic generation using a waveguide of periodically poled lithium niobate (NTT Electronics Corp.), following the amplification of the output from an interference-filter stabilized external-cavity diode laser (Optoquest Co., Ltd.) by a tapered amplifier ($\SI{1014}{nm}$). The frequency of the seed laser is stabilized with a ultralow expansion (ULE) glass cavity (ATF-6010-4, Advanced Thin Films), where the cavity length and the finesse are $\SI{10}{cm}$ and $\sim 300,000$, respectively. The laser linewidth is measured to be 20.6(6) Hz (full-width-at-half-maximum). The laser power is $\SI{25}{mW}$ at the chamber.

We perform plane-selective excitation to the $\tPt$ state in a magnetic field gradient of $\SI{20.5}{G/cm}$. To define the quantization axis along the $z$-direction, a bias magnetic field of $\SI{0.9}{G}$ is applied in the $z$-direction using a coil different from the gradient coil for MOT. The $\tPt$ excitation laser is irradiated in the horizontal plane with $\sigma_{\pm}$-polarization and tuned to the resonance of the $\ket{0}\leftrightarrow\ket{\tPt,F=3/2,m_F=3/2}$ transition. In the plane-selective excitation experiment shown in Fig.~1, a square pulse is irradiated to the atoms for $\SI{5}{ms}$, resulting in the excitation fidelity of $44.9(1.8)\si{\%}$, $53.2(1.7)\si{\%}$, and $51.4(1.8)\si{\%}$ for Plane 1, 2, and 3, respectively. In the plane-selective control experiment shown in Fig.~2, we perform the frequency sweep of a HS1 pulse from the red sidebands toward the blue sidebands by $\pm\SI{30}{kHz}$ around the resonance to improve the excitation fidelity, resulting in excitation fractions of $77.7(1.7)\si{\%}$, $83.2(2.1)\si{\%}$, and $82.4(2.0)\si{\%}$ for Planes 1, 2, and 3, respectively. 
As for the plane-selective operation performed in the experimental sequence shown in Figs.~2(e) and S1(d), the frequency of the $\SI{1014}{nm}$ laser is shifted by a fiber electro-optic modulator (EOM) to tune the laser frequency to the resonance of each plane. Since this fiber EOM is used to adjust the locking frequency of the ULE cavity, we wait $\SI{10}{ms}$ after switching.

\textbf{\textit{Repumping the metastable state.}} {\bf \textemdash}
Repumping from the $\tPt$ state to the ground state is achieved via the (6s7s)~$\tSo$ state. Two repump lasers at the wavelengths of $\SI{770}{nm}$ and $\SI{649}{nm}$ co-propagate along the same beam path as the $\tPt$ excitation laser. The $\SI{770}{nm}$ laser frequency is tuned to the $\ket{\tPt,F=3/2}\leftrightarrow\ket{\tSo,F=1/2}$ transition~\cite{Hao2023} and irradiated with an intensity of $\SI{3}{W/cm^2}$, while the $\SI{649}{nm}$ laser frequency is tuned to the $\ket{\tPz}\leftrightarrow\ket{\tSo,F=3/2}$ transition~\cite{Hao2023} and irradiated with an intensity of $\SI{0.8}{W/cm^2}$. The repumping laser is irradiated for a sufficiently long time to ensure high repumping fidelity, resulting in repumping fidelities of $97.3(3.4)\si{\%}$, $98.2(4.2)\si{\%}$, and $92.4(3.7)\si{\%}$ for Planes 1, 2, and 3, respectively, with a pulse width of $\SI{5}{ms}$. The repumping fidelity used in this analysis is corrected as described in Section S3. For Planes 1 and 2, we obtain sufficiently high fidelities within the error bars. The fidelity for Plane 3 is slightly lower than those for Planes 1 and 2, possibly due to non-uniform illumination of the repumping beam over the entire array, which can be improved by optimizing the optical system. Furthermore, we note that the 5~ms repumping time is not optimized. Future optimization could potentially enable repumping to the ground state in a few $\si{\micro s}$, similar to Ref.~\cite{Ma2023}.

\section{S.2 Z position dependence of tweezer trap parameters}
In order to know the scalability of our present 3D system, we investigate the $z$-position dependence of tweezer trap conditions of this system by using a 2D $4\times5$ square array at various $z$-positions (Fig.~S3).
In this investigation, as shown in the top panel (red) of the Fig.~S3, we homogenize the trap depths at various $z$-positions ranging from $-\SI{100}{\micro m}$ to $\SI{100}{\micro m}$ around the center along the $z$-direction by adjusting the tweezer power.
Note that, for this homogenization process, we utilize the spectroscopy of a DLS, which is a good measure for the trap depth, between the $\ket{0}\leftrightarrow \ket{\tPo, F=3/2, m_F=\pm3/2}$ transition with the DLS set to $\SI{20.4}{MHz}$ at each $z$-position.
The middle (green) panel in Fig.~S3 shows the $z$-position dependence of the required tweezer power, indicating that, as the distance from the center increases, the required power also increases to maintain a uniform trap depth.
To obtain the information on the beam waist, which is another important quantity to characterize the tweezer trap condition, we perform Raman sideband spectroscopy of the nuclear spin states in the ground state at each $z$-position.
The result of this measurement is shown in the bottom (blue) panel of Fig.~S3, indicating also the increase of the trap beam waist as the increase of the distance from the center.
While this variation of the trap beam waist is qualitatively consistent with the behavior of the required tweezer power, we find no quantitative agreement and need to investigate other technical details such as the diffraction efficiency of the SLM.

While the aforementioned power concern exists, our investigations of the trap conditions along the $z$-direction reveal that the regions within $\pm\SI{100}{\micro m}$ are suitable for tweezer experiments in our system with an objective lens with the field of view of $\SI{200}{\micro m}$, which could prepare $> 33, 000$ sites, assuming $\SI{5}{\micro m}$ interplane distance along the $z$-direction and $\SI{5}{\micro m}$ site distance in the $xy$ plane.
In the future, SLMs with smaller pixel sizes and larger display areas could enable the generation of such a large size of tweezer arrays with a lower laser power.

    \begin{figure}[tbp]
        \centering
        \includegraphics[width=\linewidth]{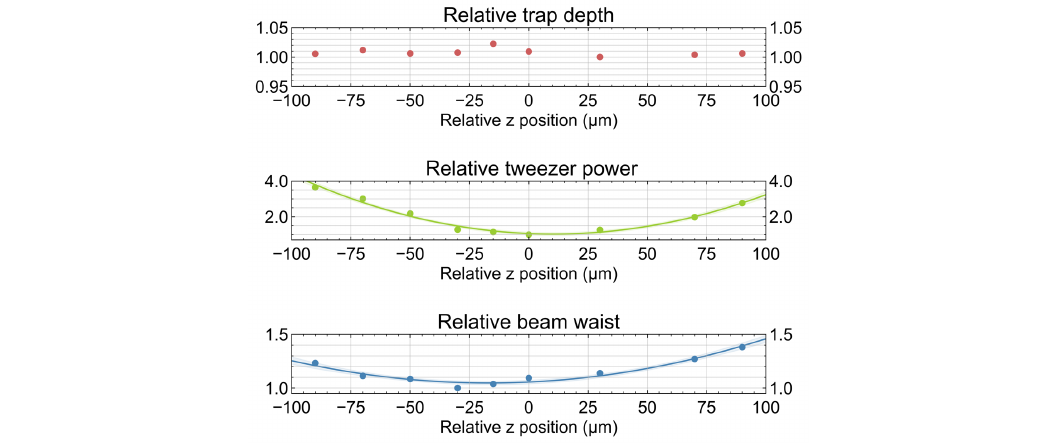}
        \caption{
        Z position dependence of tweezer trap conditions. A 4$\times$5 square lattice is used for these measurements. (Top) The trap depth is measured as the magnitude of the DLS of $\ket{0}\leftrightarrow\ket{\tPo, F=3/2, m_F=\pm3/2}$ transition, where the tweezer power is adjusted so that the average of the DLS should be 20.4 MHz. The data in the top panel are plotted as a relative ratio to the average DLS value. (Middle) The green dots indicate the required tweezer power. (Bottom) The blue dots represent the beam waist at each $z$-position measured by the Raman sideband spectroscopy. The data in the middle and bottom panel are a relative ratio to their respective minimum values. The solid lines in the middle and bottom panels are the fit lines with the quadratic curve and the shaded areas are the $1\sigma$-confidence intervals.}
        \label{fig:trappable_region}
    \end{figure}

\section{S.3 Characterization of the Plane-selective control errors}
\textbf{\textit{$\tPt$ excitation fidelity.}} {\bf \textemdash} Here, we describe a method to estimate the $\tPt$ excitation fidelity $P_{3\mathrm{P}2}$ from experimental data obtained by two sequences as follows: 

    \begin{enumerate}
        \renewcommand{\labelenumi}{\Alph{enumi})}
        \item $\tPt$ excitation fraction measurement with a repuming beam after the irradiation of a pushout beam (Fig.~1(e)),
        \item $\tPt$ excitation fraction measurement with a repuming beam alone (Fig.~1(e) with no pushout beam).
    \end{enumerate}

The two sequences are common in that only one $\tPt$ excitation and repumping process is involved. The difference is that the former is sensitive only for the atoms returned to the ground state after the $\tPt$ state excitation, and the latter also the atoms which are not excited.
More specifically, the former experimental sequence is the same as that in Fig.~1(e). From this measurement, we obtain the raw $\tPt$ excitation fraction $A^{(data)}$. The latter experimental sequence is almost the same as that in Fig.~1(e), but it does not include the pushout beam for the state-selective imaging. From this measurement, we extract the survival fraction $B^{(data)}$.
These quantities are related with each other as
\begin{equation}
    B^{(data)} = P_s(1-P_{3\mathrm{P}2})+A^{(data)}, \tag{S2}
\end{equation}
where $P_s$ represents the survival probability of the atoms after the imaging of the atoms in the ground state.
Note that we assume the ionization loss from the $\tPt$ state during plane-selective controls is small enough in the shallow trap depth of $\SI{50}{\micro K}$ (see Fig.~S4), and the ionization loss events mainly occur after ramping up the tweezer depth to $\SI{1.3}{mK}$ for imaging, whose contributions are common for $A^{(data)}$ and $B^{(data)}$. From Eq.~(S2), we obtain,
\begin{equation}
    P_{3\mathrm{P}2} = 1 - (B^{(data)} - A^{(data)})/P_s. \tag{S3}
\end{equation}
The plotted data in Fig.~1(f) and the $\tPt$ excitation fidelity values in section S.1 and the main text are corrected by the finite survival probability $P_s$ according to Eq.~(S3).

\vskip\baselineskip
\textbf{\textit{Repumping fidelity.}} {\bf \textemdash} 
In addition to the above measurements used to determine the $\tPt$ excitation fidelity, we perform an additional measurement to determine the repumping fidelity $P_r$ as follows:
\begin{enumerate}
        \renewcommand{\labelenumi}{\Alph{enumi})}
        \setcounter{enumi}{2}
        \item $\tPt$ excitation fraction measurement without a repuming beam and with a pushout beam.
\end{enumerate}
This sequence is almost the same as that in Fig.~1(e), but it does not include the repumping beam pulse. This measurement is sensitive for the atoms that decay from the $\tPt$ state to the ground state in a deep tweezer trap, and the raw leakage fraction $C^{(data)}$ is obtained according to the following relation:
\begin{align}
    C^{(data)} &= P_sP_{3\mathrm{P}2}f_{3\mathrm{P}2\rightarrow1\mathrm{S}0}, \tag{S4}
\end{align} 
where $f_{3\mathrm{P}2\rightarrow1\mathrm{S}0}$ is the leakage fraction from the $\tPt$ state to the ground state. Since $A^{(data)}$ can be written as $A^{(data)} = P_sP_{3\mathrm{P}2}\qty{P_r + (1-P_r)f_{3\mathrm{P}2\rightarrow1\mathrm{S}0}}$, we obtain the repumping fidelity $P_r$ as the following relation,
\begin{equation}
    P_r = \frac{1}{P_sP_{3\mathrm{P}2} - C^{(data)}}\qty(A^{(data)} - C^{(data)}). \tag{S5}
\end{equation}
The repumping fidelity values presented in section S.1 are corrected by the finite survival probability $P_s$ and the $\tPt$ excitation fidelity $P_{3\mathrm{P}2}$ with the HS1 pulse according to Eq.~(S5).

\vskip\baselineskip
\textbf{\textit{Plane-selective initialization fidelity}} {\bf \textemdash}
Here, we detail the survival probability correction for the plane-selective initialization experiment. To determine the initialization fidelity from the experimental data obtained using the sequence shown in Fig.~2(a), we quantify the atomic loss occurring during the plane-selective initialization sequence. To measure this atomic loss, we perform additional measurements in a slightly different sequence from Fig.~2(a) in that we do not apply the $\ket{0}$ state push beam. With this modified sequence, the second imaging becomes sensitive to atoms distributed in the ground state, allowing us to measure the atomic loss caused by the plane-selective initialization. In this case, the probability $D^{(data)}$ of an atom being detected in the second imaging is given by:
    \begin{align}
        D^{(data)}(N_{cyc}) = P_s(1-P_{loss}(N_{cyc})), \tag{S6}
    \end{align}
where $N_{cyc}$ is the number of $\tPt$ pumping cycles, $P_s$ is the survival probability of imaging, and $P_{loss}$ is the probability of atomic loss due to the $\tPt$ pumping. By dividing the $\tPt$ pumping cycle dependence data $E^{(data)}(N_{cyc})$ measured with the Fig.~2(a) sequence by $D^{(data)}(N_{cyc})$, we obtain the population $P_{1}$ of state $\ket{1}$ shown in Fig.~2(b) as follows
    \begin{align}
        P_{1}(N_{cyc}) = E^{(data)}(N_{cyc})/D^{(data)}(N_{cyc}). \tag{S7}
    \end{align}

\vskip\baselineskip
\textbf{\textit{Stability of the $\tPt$ state in a tweezer.}} {\bf \textemdash}
Toward a plane-selective mid-circuit measurement, it is necessary to shelve the atoms in all non-imaged planes into a metastable state in order to isolate them from the lasers for imaging and cooling. To evaluate the ionization rate of the $\tPt$ state in $\SI{532}{nm}$ tweezers, we measure the stability of $\tPt$ state in a tweezer trap. As shown in Fig.~S4(a), the experimental sequence begins by applying a $\SI{507}{nm}$ laser and pushing out the ground-state atoms in a shallow depth of $\SI{50}{\micro K}$. The tweezer depth is then ramped up to $U_0$, and is kept for a hold time. Finally, a repumping beam is applied to transfer the atoms remaining in the $\tPt$ to the ground states, followed by imaging. Note that this measurement is not sensitive to the photon scattering of the atoms from the $\tPt$ to the ground states via the $\tPo$ state.

Figure~S4(b) shows the trap depth dependence of the $\tPt$ state atom decay rate $\Gamma_m$, which is characterized by $\beta U_0^2$. From the fits of the data, $\beta$ are determined to be $21.5(4)~\si{Hz/(mK)^2}$, which indicates the importance of working at shallower trap depths. 
The quadratic term $\beta$ could be due to two-photon ionization~\cite{Zernik1964,Bebb1966} from the $\tPt$ state which can happen because the two-photon energy of the $\SI{532}{nm}$ tweezer laser ($\sim \SI{37590}{cm^{-1}}$) exceeds the ionization limit from the $\tPt$ state ($\sim \SI{30733}{cm^{-1}}$). Similar quadratic dependence of the decay rate of a metastable state has recently been observed in the $\tPz$ state of $\fermi$ atoms with a $\SI{486.8}{nm}$ tweezer laser~\cite{Ma2023}, which is qualitatively similar to our observation. The observed value will be compared with an appropriate model~\cite{Bebb1966,Burgers2022}.

\begin{figure}[htbp]
    \centering
    \includegraphics[width=0.5\linewidth]{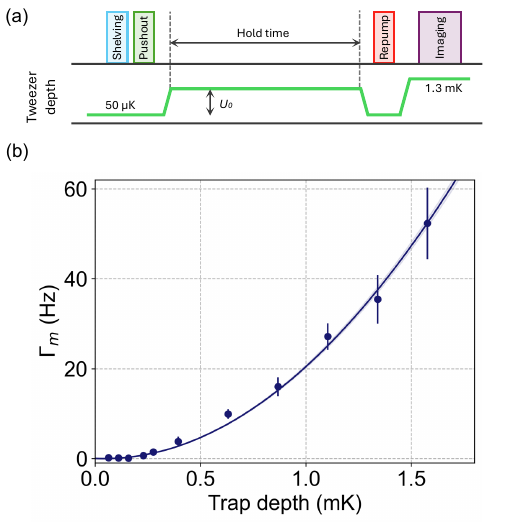}
    \caption{$\SI{532}{nm}$ tweezer light induced ionization. (a) Experimental sequence for $\tPt$ state stability measurement. (b) $\tPt$ state loss rate as a function of trap depth $U_0$. The quadratic term is $21.5(4)~\si{Hz/(mK)^2}$.}
    \label{fig:ionizatino}
\end{figure}

\vskip\baselineskip
\textbf{\textit{Simulation of the $\tPt$ excitation fidelity}} {\bf \textemdash}
To analyze the $\tPt$ excitation fidelity, we consider an atom-light interaction Hamiltonian in an 1D harmonic trap, which accounts for the motional state along the $\tPt$ excitation laser's propagation direction. With the detuning $\Delta$ from the excitation resonance, the Rabi frequency $\Omega$, and the trap frequency $\omega_r$, this Hamiltonian is given as follows,
    \begin{equation}
        \hat{H}/\hbar = (-\Delta\ket{e}\bra{e}\otimes\mathbbm{1}_{M}) + \frac{\Omega}{2}(\hat{\sigma}\otimes e^{-i\eta(\hat{a}+\hat{a}^{\dagger})} + \hat{\sigma}^{\dagger}\otimes e^{i\eta(\hat{a}+\hat{a}^{\dagger})}) + \mathbbm{1}_S\otimes\omega_r\sum_{n=0}^{N}(n+\frac12)\ket{n}\bra{n}, \tag{S8}
    \end{equation} 
where $\mathbbm{1}_S$ and $\mathbbm{1}_M$ are the identity operators for the pseudo spin space representing the optical qubit of the $\sSz$ and $\tPt$ states and the motional space, respectively. Here, $\eta$ is the Lamb-Dicke parameter, $\sigma$ and $\sigma^{\dagger}$ are the pseudo spin raising and lowering operators, and $a$ and $a^{\dagger}$ are the annihilation and creation operators for the motional degrees of freedom.
While this Hamiltonian assumes a zero differential light shift (DLS) between the ground and excited states, this assumption is justified by our experimental result that the measured trap depth dependence of DLS between $\ket{0}$ and $\ket{\tPt, F=3/2, m_F=3/2}$ is $0.58(4)~\si{Hz/(W/cm^2)}$, corresponding to DLS of $14.3(1.1)~\si{kHz}$ with the corresponding differential trap frequency of $\SI{0.19(1)}{kHz}$ at a $\SI{1.04}{MHz}(=\SI{50}{\micro K})$ trap depth.
    
To account for shot-to-shot detuning errors, we sample a particular value of detuning 1000 times from a Gaussian distribution with a FWHM of $\SI{53.1}{kHz}$ obtained from the linewidth analysis, and calculate the excited state population, and then average the populations. By determining the maximum average excited state population, we calculate the $\pi$ pulse fidelity. We note that the calculation includes motional states up to the $N' = 10$ level. All parameters used in this simulation are based on experimental parameters: Rabi frequency $\Omega = 2\pi\times \SI{10}{kHz}$, trap frequency $\omega_r = 2\pi\times \SI{28}{kHz}$, Lamb-Dicke parameter $\eta = 0.4$, and initial average vibrational quantum number $\bar{n} = 0.59$. Based on the above analysis, the simulated $\pi$ pulse fidelity is 36.3(9)\%, which roughly reproduces the experimental data of $53.2(1.7)~\si{\%}$ for Plane~2.

\vskip\baselineskip
\textbf{\textit{Simulation of the $\tPt$ excitation linewidth}} {\bf \textemdash} Here we describe the detail of the simulation of the $\tPt$ excitation linewidth. In this simulation, we consider the following factors contribute to the broadening of the excitation linewidth:
\begin{itemize}
    \item {\bf Power broadening by $\tPt$ excitation laser} \\
        Unlike typical precision measurement experiments, sufficiently high power lasers are used to observe Rabi oscillations. Therefore, when analyzing the excitation linewidth, it is necessary to consider power broadening effect of the spectrum. Given the average vibrational quantum number $\bar{n}$ and the Rabi frequency $\Omega$ of the carrier transition, the excitation spectrum including the sideband structure in a 1D harmonic potential, can be expressed as follows:
        \begin{equation}
            \rho_{ee}(\omega) = \frac{1}{\mathcal{N}}\sum_{n_g, n_e}{}e^{-\frac{\hbar\omega_r}{k_BT}\ab(n_g+1/2)} \times \frac{\ab(\frac{\Omega_{n_g, n_e}}{\Gamma})^2}{1 + \ab(\frac{2(\omega-\omega_{n_g, n_e})}{\Gamma})^2 + 2\ab(\frac{\Omega_{n_g, n_e}}{\Gamma})^2}, \tag{S9}
            \label{eq:boradening}
        \end{equation}
        where $\omega$ is a laser frequency, $\mathcal{N}$ is the normalization constant, atomic temperature $T = \hbar\omega_r/(k_B\log\ab(1+\bar{n}^{-1}))$~\cite{Wineland1987}.
        The effective linewidth of the excitation state $\Gamma = \SI{7.6(8)}{kHz}$, which is experimentally given by the decay time of the Rabi oscillation between the $\ket{0}\leftrightarrow \ket{\tPt,F=3/2,m_F=3/2}$ transition.
            
        $\Omega_{n_g, n_e}$ and $\omega_{n_g, n_e}$ represent the Rabi frequency and resonance frequency of each sideband, respectively, and are given by the following expressions~\cite{Wineland1998}:
        \begin{align}
            \Omega_{n_g,n_e} &= \Omega\exp\ab[-\eta^2/2]\sqrt{\frac{n_<!}{n_>!}}\eta^{\ab|n_e-n_g|}L_{n_<}^{\ab|n_e - n_g|}(\eta^2), \tag{S10}\\
            \omega_{n_g, n_e} &= (n_e - n_g)\omega_r, \tag{S11}
        \end{align}
        where $n_{<}~(n_{>})$ is the lesser (greater) of $n_e$ and $n_g$, and $L_n^{\alpha}$ is the generalized Laguerre polynomial. 
    
        Using our experimental parameters, the excitation linewidth due to power broadening is calculated to be $\SI{13.1}{kHz}$ for the carrier, $\SI{4.7}{kHz}$ for the 1st sidebands.
            
    \item {\bf Magnetic field inhomogeneity by the gradient coil} \\
        The anti-Helmholtz coils used to create the magnetic field gradient in the $z$-direction also produce a slight magnetic field gradient in the $xy$-plane. This can lead to variations in the Zeeman shift across different tweezer sites, contributing to the broadening of the excitation linewidth. When the center of the magnetic field is taken as the origin, the magnetic field at a tweezer site $(x_0, y_0, z_0)$ is given by:
        \begin{equation}
            \ab|B| = \sqrt{b^2\ab(x-x_0)^2/4 + b^2(y-y_0)^2/4 + \ab(b\ab(z-z_0) + B_0)^2}, \tag{S12}
            \label{eq:mag}
        \end{equation}
        where $B_0$ is the magnitude of the bias magnetic field in the $z$-direction and $b$ is a magnetic field gradient coefficient. In our experiment, $B_0 = \SI{0.9}{G}$ and $b = \SI{20.5}{G/cm}$.
        As is obvious from Eq.~(\ref{eq:mag}), the Zeeman shift is minimized when the center of the magnetic field and the center of the tweezer array coincide. Prior to the plane-selective manipulation experiments, we optimize the horizontal compensation magnetic field to minimize the $\tPt$ resonance frequency, ensuring this alignment. We assume that the horizontal position of the tweezer array and the center of the magnetic field nearly coincide by this calibration, and thus use $x_0 = y_0 = 0$ in this analysis. For the $z$-direction, we determined the displacement between the center of the tweezer array and the center of the magnetic field to be 1.2 mm by measuring the resonance frequency shift with and without the magnetic field gradient after the above calibration. Therefore, we use $z_0 = \SI{1.2}{mm}$ in this analysis. Finally, we calculate the Zeeman shift for each site using Eq.~(\ref{eq:mag}), and determine the excitation linewidth broadening due to magnetic field inhomogeneity by calculating the FWHM of their standard deviation, which we found to be 0.3~kHz.
            
    \item {\bf Magnetic field fluctuation by the gradient coil} \\
        Since the $\ket{\tPt,F=3/2, m_F=3/2}$ state has a Zeeman shift with a magnitude of $\SI{3.75}{MHz/G}$, the excitation linewidth is highly sensitive to magnetic field fluctuations. The ripple noise from the current source (PAN 35-30A, KIKUSUI Electronics Corp.) for the coils generating the magnetic field gradient has a relative standard deviation of 0.3~\%. Therefore, the linewidth broadening caused by the current ripple noise is determined to be FWHM = 51.4~kHz, which is a major factor in broadening the excitation linewidth.
        
        To confirm that the ripple noise is indeed the dominant source, we measure the $\tPt$ excitation linewidth without applying the magnetic field gradient. As a result, we obtain the measured excitation linewidth of FWHM = 2.6(4)~kHz, which is much narrower compared to those obtained when the magnetic field gradient is applied. The observation of a narrower linewidth without the magnetic field gradient supports the hypothesis that the ripple noise is the primary source (strictly speaking, the magnetic field inhomogeneity conditions also change with or without the magnetic field gradient, but based on the above analysis, the inhomogeneity effect is not dominant). 
    
    \item {\bf Stray magnetic field fluctuation} \\
        In addition to the gradient coil, stray magnetic field fluctuations can also broaden the excitation linewidth. 
        To evaluate this effect, we analyze the 2.6(4)~kHz excitation linewidth of the $\ket{\tPt, F=3/2, m_F=3/2}$ state when no magnetic field gradient is applied. Specifically, we estimate the linewidth broadening due to trap depth inhomogeneity and trap depth fluctuation during this spectroscopic measurement. By calculating the remaining linewidth broadening, the systematic uncertainty caused by stray magnetic field fluctuations is determined. Details regarding trap depth inhomogeneity and trap depth fluctuation are provided later. We note that the $\tPt$ excitation laser power is kept sufficiently low during this spectroscopic measurement to avoid power broadening.
    
        The experimental conditions for this evaluation, conducted without the applied magnetic field gradient, differ from those used in the plane-selective manipulation experiment described in the main text. The experimental parameters used for the evaluation are as follows: DLS = 14.3~kHz, trap depth inhomogeneity = 0.98~\%, trap depth fluctuation = 0.98~\%, and average vibrational quantum number = 0.19. These parameters correspond to linewidth broadenings of 0.3~kHz due to trap depth inhomogeneity, 0.3~kHz due to trap depth fluctuation. Thus, the excitation linewidth due to stray magnetic fields is estimated to be 2.5~kHz.
    
    \item {\bf Trap depth inhomogeneity} \\
        Since 532~nm is not perfectly a magic wavelength for the $\sSz \leftrightarrow \tPt$ transition, the inhomogeneity of the trap depths among the tweezers contribute to the slight broadening of the $\tPt$ excitation linewidth. We use the trap depth inhomogeneity value 0.5\%, obtained from the $\sSz \leftrightarrow \tPo$ excitation spectroscopy described in Sec.~S1 of SM. Using a Gaussian distribution with a mean value of 14.3 kHz and a relative standard deviation of 0.5\%, we calculate the FWHM of this distribution. This allows us to determine the excitation linewidth broadened due to the trap depth inhomogeneity, which we found to be 0.2~kHz.
    
    \item {\bf Trap depth fluctuation} \\
        Power fluctuations in the tweezer beam can also contribute to the broadening of the excitation linewidth. The tweezer power is stabilized by an AOM upstream of the SLM, resulting in an intensity fluctuation with a relative standard deviation of 0.98\%. By performing a calculation similar to that for trap depth inhomogeneity, the excitation linewidth broadening due to the tweezer depth fluctuation is estimated to be 0.3~kHz.
        
\end{itemize}
    
The total excitation linewidth can be obtained by calculating the quadrature sum of the individual linewidths determined for each factor. Table~S\ref{table:error_budget} summarizes the results of our analysis. This analysis yields linewidths of 53.1~kHz and 51.7~kHz for the carrier and 1st sideband transitions, respectively. To compare these numerical results with the experimental data, we plot the Lorentzian with a triple peak having these FWHMs, as shown by the solid line in Fig.~1(f). For clarity, the peak heights are normalized to the maximum value of the data points.

\vskip\baselineskip
\textbf{\textit{Error budgets of the $\tPt$ excitation}} {\bf \textemdash}
Using the excitation linewidths estimated in the above discussion, we construct a detailed error budget for the excitation fidelity. We treat each systematic linewidth broadening as a shot-to-shot detuning error and calculate the excitation fidelity by averaging 1000 samples using the method described in "Simulation of the $\tPt$ excitation fidelity." The results are summarized in Table~S\ref{table:error_budget}. This analysis shows that the magnetic field fluctuation is the limiting factor for the excitation fidelity.
    \begin{table}[h]
        \caption{Systematic broadening for the $\tPt$ excitation spectroscopy and error budget for the $\pi$-pulse excitation. The values in parentheses for the $\pi$-pulse infidelity represent the standard error of the infidelity when 1000 detunings are sampled from a Gaussian distribution with a FWHM determined by the broadening.}
        \centering
        \begin{tabular}{lcc}
            \toprule
             Error source & Broadening (kHz) & $\pi$-pulse infidelity (\%) \\
             \hline
             Power broadening & \begin{tabular}{lc} Carrier & 13.1 \\ 1st sideband & 4.7 \end{tabular} & -- \\
             Magnetic inhomogeneity &  0.3 & 0.019(1) \\
             Magnetic fluctuation \par by the gradient coil &  51.4 & 56.3(9) \\
             Stray magnetic field fluctuation & 2.5 & 1.31(6) \\
             Trap depth inhomogeneity &  0.2 & 0.009(0) \\
             Trap depth fluctuation & 0.3 & 0.019(1) \\
             Motional dephasing & -- & 6.05 \\
             \hline
             Total estimate & \begin{tabular}{lc} Carrier & 53.1 \\ 1st sideband & 51.7 \end{tabular} & 63.7(9) \\
                 
             \hline
             \hline
             \label{table:error_budget}
        \end{tabular}
    \end{table}

\newpage
\textbf{\textit{Toward high-fidelity $\tPt$ excitation}} {\bf \textemdash} As shown in the previous section, our current fidelity is not limited by a fundamental origin, but rather by technical issues.
Here, we discuss strategies for improving the $\tPt$ excitation fidelity, using the aforementioned simulation. Since the shot-to-shot detuning error is currently the dominant error (See Table~S\ref{table:error_budget}), we investigate the dependence of the fidelity on the excitation FWHM linewidth $\delta$. Furthermore, as the shot-to-shot detuning error can be suppressed by increasing the Rabi frequency $\Omega$, we also evaluate the dependence of the fidelity on the Rabi frequency.
For the simulation, we treat the linewidth $\delta$ and Rabi frequency $\Omega$ as free parameters, while the trap frequency $\omega_r$, the Lamb-Dicke parameter $\eta$, and the mean phonon number $\bar{n}_{xy}$ are fixed to experimental values of $2\pi\times\SI{28}{kHz}$, $0.4$ and $0.2$, respectively. The simulation results are shown in Fig.~S5. If we could suppress the excitation linewidth to FWHM of $\SI{2.5}{kHz}$ and achieve a Rabi frequency of 200 kHz, we expect to obtain an excitation fidelity of over 99.5\%, which is the error threshold of the surface code~\cite{Fowler2012}. The target values for excitation linewidth and Rabi frequency to achieve this error threshold could be attained by a 20-fold reduction in the current noise of the gradient coil and by a 20-fold increase in the Rabi frequency through the use of commercially available high-power lasers with a smaller beam diameter.

\begin{figure}[h]
    \centering
    \includegraphics[width=0.6\linewidth]{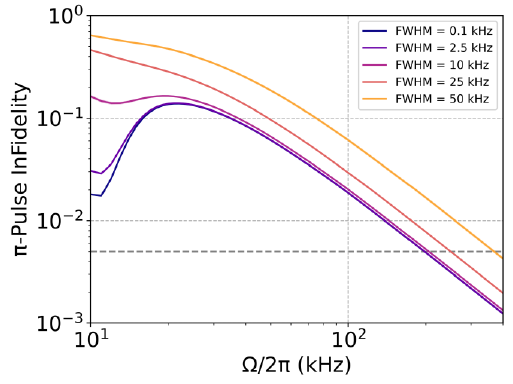}
    \caption{Shot-to-shot detuning error effect. We simulate the $\pi$ pulse fidelity of the clock excitation using our experimental parameters ($\bar{n}_{xy} = 0.2$ and a trap frequency of $2\pi\times\SI{28}{kHz}$). As the Rabi frequency increases, the excitation infidelity decreases as a general tendency with the exception for the peaks around 20 kHz which are attributed to the motional dephasing effect.}
    \label{fig:fidelity}
\end{figure}

\vskip\baselineskip
\textbf{\textit{Crosstalk characterization}} {\bf \textemdash}
To realize MHz-scale Rydberg interactions in 3D tweezer arrays, as achieved in 2D tweezers, the interplane distance should be reduced from the current $\SI{30}{\micro m}$ preferably to 2--3~$\si{\micro m}$ like in state-of-the-art two-qubit gate experiments~\cite{peper2024, Muniz2024}.
To evaluate the minimum achievable interplane distance in a 3D optical tweezer array, we investigate the interplane distance dependence of the crosstalk error in plane-selective excitation. In our simulations, we consider various effects such as the sideband structure~\cite{Wineland1979} of a trap depth of $\SI{50}{\micro K}$ during the $\tPt$ excitation (Lamb-Dicke parameter is 0.4), the residual DLS between the $\sSz \leftrightarrow \tPt$ states, the fluctuations in the Zeeman shift of the $\ket{\tPt, F=3/2, m_F=3/2}$ state caused by magnetic field fluctuations, and power broadening by the excitation laser. We note that this simulation method is also used to simulate the $\tPt$ excitation linewidth in Fig.~1(f). By using the simulated spectra, we evaluate the overlap of the spectra of different planes. Here we assume an 11-plane tweeer array with a site separation of $\SI{4}{\micro m}$ in the $xy$ plane, a trap depth inhomogeneity of 0.5\%, a bias-$z$ magnetic field of $\SI{500}{G}$, a magnetic field gradient of $\SI{300}{G/cm}$, a mean phonon occupation number of 0.2 (horizontal direction), an excitation Rabi frequency of $2\pi\times \SI{1}{kHz}$. To calculate the power broadening, we use the spontaneous emission rate of $2\pi\times \SI{14.6}{mHz}$ for the $\tPt$ state~\cite{Porsev2004}.

Figure~S6(a) shows the interplane distance dependence of the crosstalk error for the plane-selective $\tPt$ excitation in an 3D tweezer array with 11 planes. Here, the crosstalk error is defined as the sum of the false excitation probability of all non-target planes, where a $\pi$ pulse is applied at the resonant frequency of a target plane. 
The blue and orange lines in Fig.~S6(a) represent the crosstalk error with the magnetic field fluctuations suppressed to $\SI{100}{\micro G}$ and $\SI{1}{mG}$, respectively, from approximately $\SI{5}{mG}$ observed in the current setup. Solid and dashed curves represent the crosstalk errors for 40$\times$40 and 4$\times$4 tweezer arrays per plane, respectively. With the magnetic field fluctuation of $\SI{100}{\micro G}$, a crosstalk error below $\SI{0.1}{\%}$ can be achieved with interatomic distances greater than $2\si{\micro m}$ even for a large array of 40$\times$40$\times$11 sites, enabling the sufficiently plane-resolved excitation as shown in Fig.~S6(b).

\begin{figure}[h]
    \centering
    \includegraphics[width=0.6\linewidth]{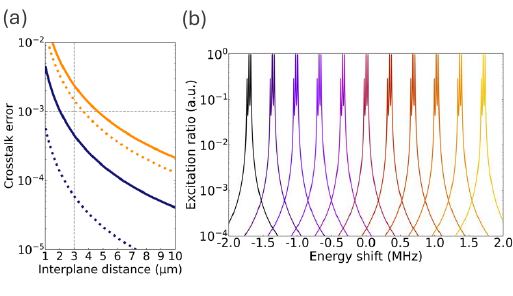}
    \caption{Crosstalk characterization for the plane-selective excitation. (a) Interplane distance dependence of the crosstalk error. Increasing the number of horizontal sites in each plane from $4\times4$ (dashed curves) to $40\times40$ (solid curves) leads to increased crosstalk errors due to the horizontal magnetic field gradient. In the case of the $40\times40$ sites, a crosstalk error of $0.1\si{\%}$ (dashed horizontal line of gray) could be achieved with an interplane distance greater than $\SI{2}{\micro m}$. The blue and orange lines represent the crosstalk error with the magnetic field fluctuations of $\SI{100}{\micro G}$ and $\SI{1}{mG}$, respectively. (b) Simulated excitation spectrum of the $\tPt$ state for the blue solid curve in (a) with the interplane distance of $\SI{3}{\micro m}$ (dashed vertical line of gray in (a)). The crosstalk of each plane at the carrier resonant frequency is less than 0.1\%. Note that the resolved sideband structure is also visible.
    }
    \label{fig:shorterdis}
\end{figure}

\newpage
%